\documentclass[]{article-hermes}
\pdfoutput=1
\usepackage{multirow}
\usepackage{colortbl}
\usepackage{amsmath}
\usepackage{subfigure}
\usepackage{textcomp}

\proceedings{Van Tien NGUYEN}

\title[Similarité entre termes et concepts]{Mesure de la similarité \\entre termes et labels de concepts ontologiques}

\author{ Van Tien NGUYEN\fup{*} \andauthor Christian SALLABERRY\fup{*} \andauthor Mauro GAIO\fup{*}}
\address{%
\fup{*} Laboratoire LIUPPA \\ BP-1155, 64013 PAU Université Cedex\\[3pt] prénom.nom@univ-pau.fr\\[6pt] }

\resume{
	Nous proposons dans cet article une méthode permettant de mesurer la similarité entre des termes et des concepts ontologiques. Notre métrique permet de prendre en compte les mots proches communs aux deux chaînes de caractères à comparer mais également d'autres caractéristiques telles que la position des mots dans ces chaînes, ou encore, le nombre d'opérations de suppression, d'insertion ou de remplacement de mots nécessaire à la construction d'une des deux chaînes à partir de l'autre. La méthode proposée a ensuite été utilisée pour déterminer les concepts ontologiques équivalents aux termes qui qualifient des toponymes. Dans un contexte de recherche d'information géographique, ceci a pour but de typer des toponymes.
}

\abstract{
We propose in this paper a method for measuring the similarity between ontological concepts and terms. Our metric can take into account not only the common words  of two strings to compare but also other features such as the position of the words in these strings, or the number of deletion, insertion or replacement of words required for the construction of one of the two strings from each other.
The proposed method was then used to determine the ontological concepts which are equivalent to the terms that qualify toponymes. It aims to find the topographical type of the toponyme.
}

\motscles{
mesure de similarité, métrique hybride
}


\begin{document}

\maketitlepage

\section{Introduction}
\label{begin_document}
L'appariement d'ontologies \cite{euzenat2007} vise à trouver des correspondances entre entités d'ontologies différentes. Ces correspondances reposent notamment sur l'existence de propriétés similaires : des relations d'équivalence, de conséquence, de subsomption entre entités, etc. Classiquement, les entités à comparer sont des classes d'une ontologie, ses propriétés et ses individus.


Le résultat du processus d'appariement appelé alignement est l'ensemble des correspondances entre deux ontologies. Notre proposition peut-être considérée comme une étape préalable à tout processus d'appariement.  Il s'agit de la comparaison de chaînes de caractères. Étape incontournable  lorsqu'il est nécessaire de comparer des entités ontologiques et que celles-ci sont accompagnées de labels constitués de termes permettant d'expliciter leur sens.

Ainsi, notre travail vise à établir s'il existe des relations d'équivalence entre chaque terme label d'une entité d'une ontologie et chaque terme label de chaque entité d'une autre ontologie. Ou alors,  s'il existe des relations d'équivalence entre chaque terme d'un lexique et chaque terme label de chaque entité d'une ontologie.
Le problème est donc celui de la comparaison de deux chaînes de caractères constituées par la paire que nous nommerons par convenance (terme, label). Ces chaînes de caractères ont comme caractéristique fréquente d'être composées de plusieurs mots. Considérons quelques paires (terme, label) : (\textit {chemin de fer touristique, voie ferrée touristique}), (\textit {centre de formation professionnelle des adultes, centre de formation des adultes}), (\textit {nation, haras national}), (\textit {poste de radio, bureau de poste}), etc. Quel est le score de similarité pour chaque paire ? Une réponse à cette question est portée par les métriques de comparaison de chaînes de caractères, que nous qualifierons par convenance de métriques de chaînes.

Nous avons choisi comme cadre expérimental d'utiliser l'ontologie géographique \cite{Mustiere11}  créée dans le cadre du projet ANR GéOnto que nous exploitons à des fins d'indexation spatiale de documents textuels \cite{Joliveau11}. Dans ce cadre, le processus mis au point pour indexer une entité nommée spatiale nécessite de lui attribuer un type, comme par exemple, \textit{hydronyme, horonyme, voies de communication, lieu dit habité}, etc.). Si on prend l'exemple du syntagme  \textit{chemin de fer touristique d'Artouste}, l'entité nommée spatiale est \textit{Artouste} et le type associé est \textit{voie de communication}. Ce typage est supporté par l'algorithme de mesure de similarité proposé dans cet article. Il permet d'apparier le terme \textit{chemin de fer touristique}, extrait du document à indexer et le label \textit{voie ferrée touristique} de l'ontologie géographique.

Le document est organisé comme suit. Dans la section suivante, nous discutons du rôle de la comparaison de chaînes de caractères dans les techniques d'appariement d'ontologies et, par conséquent, dans les métriques de chaîne. Dans la section 3, nous proposons une méthode qui permet de comparer des chaînes. L'expérimentation de cette méthode sera présentée et discutée dans la section 4.
\section{État de l'art}
\subsection{Le rôle de la comparaison de chaînes de caractères dans des techniques d'appariement d'ontologies}


Plusieurs techniques ont été proposées afin de résoudre le problème d'appariement d'ontologies. Euzenat et Shvaiko \cite{euzenat2007} ont fait un état de l'art approfondi relatif à ces techniques  après avoir analysé une cinquantaine de systèmes différents.
Ils distinguent des techniques de deux niveaux différents : le niveau élémentaire et le niveau structurel. Les techniques de niveau élémentaire considèrent les entités sans tenir compte de leurs relations avec d'autres entités dans une même ontologie. Les techniques de niveau structurel, quant à elles, comparent non seulement les entités mais aussi leurs relations avec d'autres entités.
Dans plusieurs systèmes d'alignement d'ontologies, les techniques élémentaires et structurelles ont été combinées afin de résoudre le problème de l'appariement d'ontologies.
En général, une ou plusieurs techniques élémentaires sont utilisées avant d'appliquer des techniques structurelles ; voici quelques exemples de systèmes d'alignement d'ontologies :
\begin{itemize}
\item COMMA \cite{Do2005} : techniques basées sur les chaînes \footnote{Il s'agit des techniques de niveau élémentaire. Ces techniques considèrent les entités des ontologies comme des chaînes de caractères. L'idée principale est que plus les chaînes de caractères (les labels des concepts, par exemple) sont similaires,
plus les concepts correspondants peuvent être considérés comme proches.}, techniques linguistiques \footnote{Il s'agit également des techniques de niveau élémentaire. Ces techniques utilisent différents outils de TAL afin d'exploiter les propriétés morphologiques (le lemme, la catégorie grammaticale, par exemple) des mots dans les chaînes à comparer.}  et techniques basées sur les graphes \footnote{Il s'agit des techniques de niveau structurel. Ces techniques représentent les ontologies par des graphes conceptuels.} ;	 
\item S-Match \cite{Giunchiglia2003} : techniques basées sur les chaînes, techniques linguistiques, techniques basées sur les ressources linguistiques \footnote{Il s'agit des techniques de niveau élémentaire. Elles utilisent de telles ressources afin d'exploiter les relations de synonymie ou d'hyponymie.} et techniques basées sur les graphes ;
\item Taxomap \cite{Hamdi2008}: techniques basées sur des chaînes, techniques linguistiques et techniques structurelles ;
\item ASCO \cite{bach2006} : techniques basées sur des chaînes et techniques structurelles.
\end{itemize}

Comme nous pouvons le constater, les travaux abordés ci-dessus ont montré que l'étape qui consiste à comparer les chaînes de caractères qui représentent des entités des ontologies (concepts, labels, relation) est une étape préalable et incontournable dans le processus d'appariement d'ontologies. Les chaînes de caractères sont comparées à l'aide des métriques de chaînes que nous présentons dans la section \ref{section_metriques_de_chaine} ci-après.



\subsection{Les métriques de chaînes et notre problématique}
\label{section_metriques_de_chaine}
\begin{figure}[h]
\centering
\includegraphics[width=0.8\textwidth]{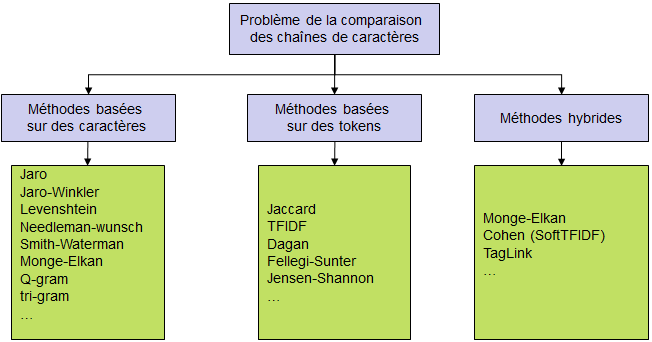}
\caption{Classification des métriques de chaîne}
\label{fig_classification_metriques_de_chaine}
\end{figure}

Les termes et les labels que nous avons besoins de comparer sont représentés normalement par les chaînes de caractères qui peuvent se composer d'un mot ou d'un groupe de mots. Comme montré par la figure \ref{fig_classification_metriques_de_chaine}, les métriques de chaîne peuvent être classées en 3 catégories principales : les méthodes basées sur des caractères, les méthodes basées sur des tokens et les méthodes hybrides.

Les métriques basées sur des caractères considèrent les chaînes comme une séquence de caractères. En conséquence, la similarité entre deux chaînes est déterminée par des caractères communs et la position de ces caractères dans les chaînes (Jaro \cite{Jaro1989}, Jaro-Winkler \cite{Winkler1999}) ou par le nombre d'opérations ($suppression, insertion, remplacement$) nécessaires pour construire une chaîne à partir de l'autre (Levenshtein \cite{Levenshtein1966}, Needleman-wunsch \cite{Needleman1970}, Smith-Waterman \cite{Smith1981}). L'inconvénient principal des ces métriques est la non distinction des mots lorsqu'une chaîne en comporte plusieurs.

Le principe de la métrique Levenshtein a été repris dans plusieurs travaux. Ceux de \cite{Zhang2010}, \cite{Wang2010} sont basés sur des structures de type B-arbre et proposent des techniques permettant d'améliorer le calcul lors de la comparaison des chaînes de caractères. Toutefois, comme dans le cas de la métrique Levenshtein, ces métriques ne traitent pas l'ordre des tokens dans la chaîne de caractères, bien que l'ordre des caractères est lui pris en compte. Il existe également des travaux, tel que I\_Sub de \cite{StoilosSK05} qui mesurent la similarité par l'intermédiaire de la sous-séquence commune la plus longue. Dans \cite{Gorbenko2012}, ce dernier problème est formalisé sous la forme d'un problème SAT. Notons que, comme pour les autres méthodes, seul l'ordre des caractères est considéré lors de la détermination de la sous-séquence commune la plus longue.

Les méthodes basées sur des tokens (TFIDF \cite{Cohen2003}, Jaccard \footnote{http://en.wikipedia.org/wiki/Jaccard\_similarity}, \cite{Marios2010}) considèrent une chaîne comme un ensemble de tokens. Un token est une sous-chaîne de caractères délimitée par des caractères spécifiques (\textit{espaces, tirets,~\ldots}).
La métrique TFIDF, comme son nom l'indique, réutilise la technique TFIDF de recherche d'information en considérant le corpus à interroger et la requête comme deux chaînes de caractères à comparer. En conséquence, la similarité entre deux chaînes sera déterminée par les tokens communs et leur fréquence dans chaque chaîne. La métrique de Jaccard, quant à elle, détermine le rapport entre le nombre de tokens communs et le nombre total de tokens distincts. Ces méthodes ne mettent pas en \oe uvre de métriques basées sur des caractères pour déterminer la similarité des tokens. Par exemple, pour les chaînes \textit{chemin de fer touristique} et \textit{voie ferrée touristique}, les tokens \textit{fer} et  \textit{ferrée} sont considérés comme différents.

Les méthodes hybrides (SoftTFIDF, Monge-Elkan, TagLink) proposent de combiner les deux types d'approches ci-dessus. Ces trois méthodes utilisent une métrique basée sur les caractères pour évaluer le degré de similarité de paires de tokens. En effet, la méthode SoftTFIDF \cite{Cohen2003} améliore la méthode TFIDF abordée ci-dessus en utilisant une métrique basée sur les caractères (JaroWinkler, par exemple) pour déterminer des couples de tokens similaires (les tokens \textit{fer} et \textit{ferrée} seront considérés comme identiques) avant d'appliquer la technique TFIDF. D'autre part, l'idée principale de la métrique TagLink \cite{Camacho2006} est de considérer le problème de comparaison de chaînes comme étant celui d'affectation, un problème classique de recherche opérationnelle : (i) les caractères d'un token sont comparés à ceux des autres tokens (un score de similarité est calculé pour chaque paire de tokens) ; (ii) les tokens dans une chaîne sont comparés à ceux de l'autre chaîne (un score global de similarité est calculé pour les deux chaînes). La différence principale entre cette méthode et les méthodes TFIDF et SoftTFIDF est que le score de TagLink dépend du rapport entre le nombre de caractères communs entre des tokens et le nombre total de caractères des tokens.
Dans la méthode Monge-Elkan \cite{Monge1996}, à l'aide d'une métrique basée sur des caractères (JaroWinkler, par exemple), pour chaque token de la chaîne $S_1$, on cherche le token le plus proche dans la chaîne $S_2$ et le score correspondant. Le score global de similarité entre deux chaîne $S_1$, $S_2$ correspond à la valeur moyenne de ces scores.

La caractéristique commune aux méthodes hybrides et aux méthodes basées sur les tokens est qu'il n'y a pas de prise en compte de l'ordre des tokens dans les chaînes (par exemple, le score de similarité calculé par Jaccard, TFIDF, SoftTFIDF, Monge-Elkan ou TagLink pour les chaînes \textit{piste de ski} et \textit{ski de piste} est égal~à~1).

\begin{table}[h]
\subfigure{
	\includegraphics[width=1.0\textwidth]{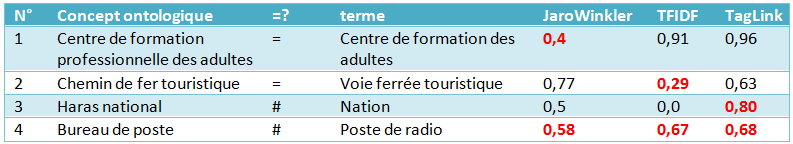}
}
\caption{Quelques exemples pour chaque famille de métriques}
\label{tableau_score_par_famille_metrique}
\end{table}

Le tableau ~\ref{tableau_score_par_famille_metrique} reporte le score obtenu par chaque famille de métriques ci-dessus pour les exemples proposés dans la section \ref{begin_document}. Les scores en rouge  illustrent les cas dans lesquels ces métriques ne marchent pas. Comme montré par ce tableau, ces métriques produisent soit un score trop faible pour les paires positives (JaroWinkler pour la paire~1, TFIDF pour la paire~2), soit un score trop élevé pour les paires négatives (TagLink pour la paire~3; les trois métriques pour la paire~4). En effet, pour la paire~1, JaroWinkler calcule une similarité faible à cause de la taille du mot \textit{professionnelle} dans la première chaîne. Ceci  implique une distance importante entre les caractères communs \og d \fg , \og e \fg , \og s \fg , \og a \fg , \og d \fg , \og u \fg , \og l \fg , \og t \fg , \og e \fg , \og s \fg{}  dans les deux chaînes : en effet, \og a \fg{}  est au rang~41 dans la première chaîne et au rang~26 dans la seconde. Pour la paire~2, TFIDF considère les tokens \textit{fer} et \textit{ferrée}  comme différents, ce qui réduit le nombre de tokens communs entre les deux chaînes. Pour la paire~3, TagLink calcule un score entre deux chaînes qui est proportionnel au score entre les tokens \textit{national} et \textit{nation}.

Puisque ces méthodes s'adaptent mal aux spécificités liées à notre problématique, nous proposons  une nouvelle métrique qui sera présentée et expérimentée ci-après.
\section{Proposition d'une métrique hybride pour la mesure de la similarité des termes}
\subsection{Formalisation de la méthode}
 Nous considérons les chaînes à comparer comme des séquences de tokens. Notre objectif est de construire une méthode qui permet de traiter les tokens avec les mêmes principes que ceux adoptés dans les méthodes basées sur les caractères telles que \cite{Jaro1989}, \cite{Levenshtein1966}, etc.


Soit $S$ l'ensemble des chaînes de caractères. Notre métrique est définie comme une fonction $\mu : S  \times  S \rightarrow \mathbf{R}$ tel que :
\begin{equation}
\begin {array}{ll}
~&0 \leq \mu(S_1, S_2) \leq 1, \forall S_1, S_2 \in S \\
~&~\\
et& \mu(S_1, S_1)=1, \forall S_1 \in S
\end {array}
\label{metrique}
\end{equation}

%

Nous souhaitons que la valeur de la fonction $\mu(S_1, S_2)$ dépende non seulement des tokens égaux (ou presque similaires) aux deux chaînes, mais dépende, également, d'autres caractéristiques de leurs tokens (comme l'ordre, ou la position des tokens dans les chaînes de caractères ; le nombre d'opérations de suppression, d'insertion ou de remplacement de tokens nécessaire à la construction d'une chaîne à partir de l'autre,~etc.).


Pour cet objectif, la valeur de la fonction $\mu(S_1, S_2)$ est calculée en deux étapes principales :

\subsubsection*{\textbf{Étape 1 - } transformation des tokens en symboles}

Chaque chaîne de caractères est considérée comme étant une liste de tokens, tel que: $S_1=\{{t_1, t_2, ...,t_n}\}\quad et\quad S_2=\{t_{n+1}, t_{n+2}, ...,t_{n+m}\}$.

Chaque token sera représenté par un symbole à l'aide de la fonction $\tau : T \rightarrow G$ avec $T=\{t_1, t_2, ...,t_{n+m}\}$  un ensemble de tokens et $G=\{\alpha_1, \alpha_2,...\}$ un ensemble de symboles prédéfinis. En conséquence, les chaînes de tokens seront représentées par des chaînes de symboles : $\tau : T^m \rightarrow G^m$ par homomorphisme de chaîne.

Dans cette étape, nous utilisons une métrique ($\mu_1$) basée sur les caractères (telle que Jaro, Levenshtein, etc) pour comparer les paires de tokens : $\mu_1 : T \times T \rightarrow \mathbf{R}$. Par conséquent, si deux tokens sont équivalents (selon la métrique et le seuil de similarité retenus) ils seront représentés par un même symbole : \\
si $\mu_1(t, t') \geq \varepsilon$ ($\varepsilon$ > 0 : un seuil prédéfini), alors $\tau(t) = \tau(t')$.

La valeur de la fonction $\tau(t), \forall t \in T$ est déterminée comme suit :

\begin{itemize}
\item Soit $T'$ l'ensemble de tokens remplacés par des symboles, initialement $T'=\emptyset$.

\item $\tau(t_1) = \alpha_1; G=G \setminus \{\alpha_1\}; T' = T' + \{t_1\} $ : le premier symbole est retiré de l'ensemble des symboles contenus dans G pour représenter le premier token $t_1$ de la chaîne $S_1$ et ce token est ajouté à $T'$.

\item $\forall t_i \in T, i > 1 :$ les symboles pour les autres tokens sont déterminés par les itérations suivantes, pour chaque itération, nous vérifions si $\exists~t' \in T'$ tel que  $\mu_1(t_i, t') \geq \varepsilon$ :
\begin{list}{+}{}
\item Si oui,  $\tau(t_i) = \tau(t') $ : parmi les tokens déjà remplacés par des symboles, on cherche le token $t'$ qui est similaire au token $t_i$. Si $t'$ existe, alors le token $t_i$ sera représenté par le symbole correspondant au token $t'$.
\item Si non, $\tau(t_i) = \alpha_x, \alpha_x \in G ; G = G \setminus \{\alpha_x\} ; T' = T' + \{t_i\} $ : un symbole est retiré de G pour remplacer le token $t_i$ et $t_i$ est ajouté à $T'$.
\end{list}
\end{itemize}
Après cette étape:
$\tau(S_1)=S'_1=\{\alpha_{i_1}\alpha_{i_2}...\alpha_{i_n}\}\;$et$ \; \tau(S_2)=S'_2=\{\alpha_{j_1}\alpha_{j_2}...\alpha_{j_m}\}$ dont les symboles peuvent être différents ou identiques.

\subsubsection*{\textbf{Étape 2 -} utilisation d'une métrique basée sur des caractères pour mesurer la similarité des chaînes de symboles}

Notons que les séquences de symboles $S'_1$ et $S'_2$ à comparer sont également des chaînes de caractères dont chaque caractère est un symbole. Dans cette étape, nous utilisons donc une deuxième métrique basée sur les caractères $\mu_2 : S \times S \rightarrow R$ pour calculer la similarité entre ces séquences de symboles.

Par conséquent, la similarité entre deux chaînes $S_1$ et $S_2$ est calculée par la fonction suivante :
$\mu(S_1, S_2)= \mu_2(\tau(S_1), \tau(S_2))=\mu_2(\alpha_{i_1}\alpha_{i_2}...\alpha_{i_n}, \alpha_{j_1}\alpha_{j_2}...\alpha_{j_m})$
%

\bigskip
En fait, notre méthode utilise deux métriques de base qui sont paramétrables : $\mu_1$ pour comparer une paire de tokens, $\mu_2$ pour comparer deux séquences de symboles. $\mu_1$ et $\mu_2$ peuvent être une même métrique ou bien deux métriques distinctes. Ainsi, chaque combinaison de métriques produit une nouvelle métrique hybride. Notre proposition correspond à une méta-méthode permettant de générer autant de méthodes distinctes que de combinaisons possibles.

C'est la métrique $\mu_2$ qui détermine les caractéristiques de nos méthodes. Par exemple si $\mu_2$ est JaroWinkler, on peut dire que notre métrique
prend en compte des tokens communs et la position des tokens dans les chaînes. Si $\mu_2$ est Levenshtein, notre métrique correspond au coût de transformation d'une chaîne vers l'autre par suppression, ajout ou remplacement des tokens.
\subsection{Illustration par deux exemples}

\subsubsection*{Exemple 1}

Considérons deux chaînes à comparer qui désignent le même concept: chaînes $S_1$= \textit{centre de formation professionnelle des adultes} , $S_2$= \textit{centre de formation des adultes}. Les paramètres d'entrée sont : $G=\{\alpha_1, \alpha_2, \alpha_3, ..., \alpha_{11}\}$ l'ensemble de symboles, $\mu_1$=JaroWinkler, $\varepsilon = 0,84$. 
Le déroulement de l'étape 1 de notre méthode est illustré par le tableau \ref{tab_deroulement_ex1}.

\begin{table}[!h]
\begin{scriptsize}
\begin{tabular}{|p{0.15cm}|p{0.15cm}|p{1.33cm}|p{1.2cm}|p{1.cm}|p{0.8cm}|p{2.3cm}|p{2.05cm}|}


\cline{3-8}

\multicolumn{2}{c|}{}&\multirow{2}{*}{Token ($t_i$)} & \multicolumn{2}{c|}{Token le plus similaire ($t'$)} & Symbole ($\tau(t_i)$) &\multirow{2}{*}{Tokens remplacés ($T'$)} & \multirow{2}{*}{Liste de symboles ($G$)} \\\cline{4-5}

\multicolumn{2}{c|}{}	&				& Token ($t'$) & Score ($\mu_1(t_i, t')$)& & & \\\cline{3-8}
\multicolumn{2}{c|}{} & & & & & & \{$\alpha_1, \alpha_2, ..., \alpha_{11}$\} \\\cline{1-8}

%
%

\multirow{6}{*}{$S_1$}&$t_1$ & centre 				& 		& 		& $\alpha_1$& \{ $t_1$\}				& \{$\alpha_2, ..., \alpha_{11}$ \}\\\cline{2-8}

& $t_2$ & de 	   			& 		& 		& $\alpha_2$& \{ $t_1, t_2$\}		&\{$\alpha_3, ..., \alpha_{11}$ \}\\\cline{2-8}

&$t_3$ & formation 		& 		& 		& $\alpha_3$& \{$t_1, t_2, t_3$\} & \{$\alpha_4, ..., \alpha_{11}$\} \\\cline{2-8}
&$t_4$ & professionnelle & 		& 		& $\alpha_4$&\{ $t_1, t_2, t_3, t_4$\} &  \{$\alpha_5, ..., \alpha_{11}$\} \\\cline{2-8}

&$t_5$ & des &			 $t_2$=de& 0,91 > $\varepsilon$& $\alpha_2$& \{ $t_1, t_2, t_3, t_4, t_5$\}&  \{$\alpha_5, ..., \alpha_{11}$\} \\\cline{2-8}

&$t_6$ & adultes 		& 		& 		&$\alpha_5$ & \{ $t_1, t_2, ..., t_6$\}& \{$\alpha_6, ..., \alpha_{11}$\} \\\hline

\multirow{5}{*}{$S_2$}&$t_7$ & centre 			& $t_1$ = centre& 1	> $\varepsilon$& $\alpha_1$& \{ $t_1, t_2, ..., t_7$\}& \{$\alpha_6, ..., \alpha_{11}$\} \\\cline{2-8}

&$t_8$ & de 		&	 $t_2$=de& 1 > $\varepsilon$&$\alpha_2$ & \{ $t_1, t_2, ..., t_8$\}& \{$\alpha_6, ..., \alpha_{11}$\} \\\cline{2-8}

&$t_9$ & formation 		& $t_3$=formation & 1 > $\varepsilon$& $\alpha_3$&\{ $t_1, t_2, ..., t_9$\} &  \{$\alpha_6, ..., \alpha_{11}$\} \\\cline{2-8}

&$t_{10}$ & des 			& $t_5$=des & 1 >  $\varepsilon$ & $\alpha_2$& \{ $t_1, t_2, ..., t_{10}$\}&  \{$\alpha_6, ..., \alpha_{11}$\} \\\cline{2-8}

&$t_{11}$ & adultes 		& $t_6$ = adultes &1 >  $\varepsilon$ & $\alpha_5$& \{ $t_1, t_2, ..., t_{11}$\}& \{$\alpha_6, ..., \alpha_{11}$\} \\\hline

\end{tabular}
\end{scriptsize}
\caption{Illustration du déroulement de l'étape 1 de notre méthode}
\label{tab_deroulement_ex1}
\end{table}

L'idée principale est de considérer une chaîne de caractères comme une séquence de tokens,  chaque token étant représenté par un symbole $\alpha_{i}$.
Dans cet exemple, la chaîne $S_1$ est composée des tokens de $t_1$ à $t_5$: \textit{ centre,  de, formation, professionnelle, des, adultes}.  Ces tokens seront représentés respectivement par les symboles $\alpha_1$, $\alpha_2$, $\alpha_3$, $\alpha_4$, $\alpha_2$, $\alpha_5$. En conséquence, $S_1$ sera représentée par une nouvelle chaîne de symboles $S'_1\:=\:\alpha_1\alpha_2\alpha_3\alpha_4\alpha_2\alpha_5$. Notons que les tokens \textit{de}  et \textit{des}  sont représentés par le même symbole ($\alpha_2$) car ces tokens sont considérés comme similaire ($\mu_1$(\textit{de}, \textit{des}) $= 0,91 > \varepsilon$).

La chaîne $S_2$ est composée les tokens allant de $t_7$ à $t_{11}$ \textit{centre, de, formation, des, adultes}.  
Comme nous pouvons le constater, les tokens identiques ou similaires sont représentés par un même symbole : par exemple, le token $t_7=$ \textit{centre} de la chaîne $S_2$ sera représenté par le symbole $\alpha_1$ car ce symbole correspond au token $t_1=$ \textit{centre} de la chaîne $S_{1}$. De même, les tokens \textit{de,  formation, des, adultes}  de la chaîne $S_2$ seront respectivement représentés par les symboles $\alpha_2$, $\alpha_3$, $\alpha_2$, $\alpha_5$. En résultat, la chaîne $S_{2}$ sera représentée par la chaîne de symboles $S'_2\:=\:\alpha_1\alpha_2\alpha_3\alpha_2\alpha_5$.

Le tableau \ref{tableau_autres_scores_par_mu2} présente  les scores obtenue par différentes métriques  $\mu_2$ sur deux exemples distincts. La deuxième colonne indique le score calculé pour les chaînes de symboles $S'_1$et $S'_2$. C'est aussi le score des chaînes $S_1$et $ S_2$.
\begin{table}[!h]
\scriptsize
\centering
\begin{tabular}{l | l l}
$\mu_2$ & score exemple 1 &   score exemple 2\\

\hline
Leveinshtein & 0,83 & 0,33\\
NeedlemanWunch & 0,83 & 0,67 \\
SmithWaterman & 0,90  & 0,33\\
MongeElkan & 0,80 & 0,33\\
Jaro & 0,94 & 0 \\
JaroWinkler & 0,96 & 0 \\
qgram & 0,67 & 0 \\

\end{tabular}
\caption{Scores obtenus par différents métriques $\mu_2$ }
\label{tableau_autres_scores_par_mu2}
\end{table}

\subsubsection*{Exemple 2}

Considérons maintenant les deux chaînes de caractères suivantes qui désignent deux concepts différents : $S_1$  = \textit{bureau de poste} et $S_2$ = \textit{poste de radio}.
Les chaînes de symboles correspondantes seront : $S'_1$ = $\alpha_1\alpha_2\alpha_3$ et $S'_2$ = $\alpha_3\alpha_2\alpha_4$.
La troisième colonne du tableau  \ref{tableau_autres_scores_par_mu2} montre les scores obtenus avec différents paramétrages de  $\mu_2$.

%
\subsection{Implémentation}
Nous avons implémenté notre méthode sous forme de module Java. Dans ce module, nous reprenons différentes métriques (basées sur les caractères) fournies par deux projets \textit{open source} SimMetrics \footnote{http ://sourceforge.net/projects/simmetrics} et SecondString \footnote{http://sourceforge.net/projects/secondstring} pour paramétrer les variables $\mu_1$ (la métrique qui compare token par token) et $\mu_2$ (la métrique qui compare les chaînes de symboles).

Le tableau \ref{tableau_mu1_mu2_epsilon} présente le code des métriques et le seuil $\varepsilon$ appliqué à la métrique $\mu_1$ au delà duquel deux tokens sont considérés comme équivalent. La valeur de $\varepsilon$ pour chaque métrique a été déterminée de manière empirique à l'aide d'une base lexicale composée de paires de tokens équivalents. Notons que nous avons implémenté la métrique Jaccard\_2 en adaptant l'algorithme de la métrique de Jaccard : ainsi, le score renvoyé par la métrique Jaccard\_2 correspond au ratio de la cardinalité des caractères communs aux deux chaînes sur la cardinalité de l'union des caractères des deux chaînes.

 On désigne désormais notre métrique \textit{Liuppa(i, j)} dont $i$ et $j$ sont respectivement le code des métriques listées dans le tableau \ref{tableau_mu1_mu2_epsilon}. Chaque paire $(i, j)$ définit une nouvelle métrique hybride. Nous avons donc 81 métriques hybrides. Ces métriques sont la combinaison deux à deux des 9 métriques présentées dans le tableau \ref{tableau_mu1_mu2_epsilon}. 

\begin{table}
\centering
\scriptsize
\begin{tabular}{ l | l | l }
Code & Métrique ($\mu_1$ ou $\mu_2$) & $\varepsilon$ (seuil pour $\mu_1$) \\
\hline
1 & JaroWinkler & 0,84 \\
2 & Levenshtein & 0,79 \\
3 & Needleman Wunch & 0,88 \\
4 & Smith Waterman & 0,83 \\
5 & Qgram & 0,60 \\
6 & Monge Elkan & 0,84 \\
7 & Jaro & 0,80 \\
8 & Jaccard\_2 & 0,80 \\
9 & I\_Sub & 0,80 \\
\end{tabular}
\caption{Paramètres de la méthode $Liuppa(i,j)$ : codes des métriques et seuils correspondants}
\label{tableau_mu1_mu2_epsilon}
\end{table}
\section{Expérimentation}
\subsection{Protocole d'expérimentation}
Notre objectif est de comparer nos métriques avec des métriques existantes.  Pour cela, nous avons repris la méthodologie d'expérimentation de Cohen \cite{Cohen2003} pour deux jeux de données différents. Selon cette démarche, l'expérimentation se fait en quatre étapes :

\begin{description}
\item [\textit{Étape 1} -] Définir le jeu de données : une liste de paires correctes et une liste de paires incorrectes sont définies manuellement par des experts (cf. section \ref{exp_sur_donnee_ontologique}) ou de manière automatique guidée par des règles \cite{Cohen2003}. Une paire est dite correcte si deux chaînes de caractères font référence au même concept : la paire (\og ville de Pau \fg , \og ville paloise \fg ) par exemple. Elle est dite incorrecte dans le cas contraire : la paire (\og Pau \fg , \og Paris \fg ) par exemple. La construction des listes de paires sera expliquée dans les sections \ref{exp_sur_donnee_ontologique}, et \ref{exp_sur_donnee_de_Cohen}.

\item [\textit{Étape 2} -] Calculer le score : le score sera calculé pour chaque paire par la métrique à évaluer.
\item [\textit{Étape 3} - ] Trier les paires : les paires sont triées en fonction de leur score de manière descendante.
\item [\textit{Étape 4} -] Calculer la précision moyenne ($avgPrecis$) à partir des résultats obtenus à l'étape 3 et des deux ensembles de paires positives et négatives de départ.
\end{description}

La précision moyenne est calculée par la formule suivante :
\begin{equation}
\label{equation_avgPrecis}
avgPrecis= \frac{1}{m} (\sum_{i=1}^{n} {x*Precision_i})
\end{equation}
où
\begin{itemize}
\item  $n$ est le nombre total de paires et $m$ celui des paires correctes ;

\item $i$ est l'ordre de la $i^{ème}$ paire dans la liste triée obtenue après l'étape~3 ;
\item $x= \left\{
			\begin{array}{rl}
				0  &\text{si la } i^{\text{ème}} \text{ paire est incorrecte}; \\
				1  & \text{si la } i^{\text{ème}} \text{ paire est correcte}.
			\end{array}
		 \right.$

\item $Precision_i = \frac{n_i}{i}$ où $n_i$ est le nombre de paires correctes avant la $i^{\text{ème}}$ paire.

\end{itemize}

%

Plus le valeur de $avgPrecis$ est grande, plus la métrique est performante.

Le tableau \ref{tableau_protocol_experimentation} illustre notre démarche d'évaluation par un exemple concret.
\begin{table}[!h]
\centering
\subfigure{
	\includegraphics[width=0.9\textwidth]{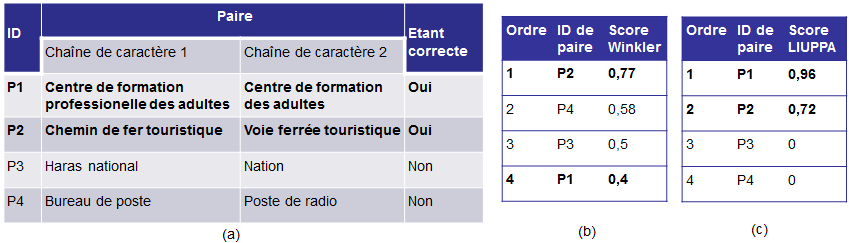}
}
\caption{Protocole d'expérimentation}
\label{tableau_protocol_experimentation}
\end{table}
Le sous-tableau \ref{tableau_protocol_experimentation}(a) décrit un jeu de données qui se compose de deux paires correctes et de deux incorrectes. Les sous-tableaux \ref{tableau_protocol_experimentation}(b) et \ref{tableau_protocol_experimentation}(c) décrivent respectivement le résultat de l'étape 3, c'est-à-dire le score des paires évalué par les deux métriques Winkler et $Liuppa(1,1)$. Notons que ces scores sont triés par ordre décroisant pour chaque métrique. La précision finale de la métrique Winkler est calculée comme suit :
$$
avgPrecis_{W}= \frac{1}{m} (\sum_{i=1}^{n} {x*Precision_i})=
\frac{1}{2}
	\left( 1.\frac{1}{1}
		+ 0.\frac{0}{2}
		+ 0.\frac{1}{3}
		+ 1.\frac{2}{4}
	\right)
	= 0,75
$$
De la même manière, nous calculons le score de la métrique $Liuppa(1,1)$ qui est : $avgPrecis_{L} = 1,0$.  Par conséquent, sur cet exemple, nous constatons que la métrique $Liuppa(1,1)$ donne les meilleurs résultats.


\subsection{Les métriques à évaluer}
Nous avons appliqué la démarche ci-dessus sur deux jeux de données différents. Les métriques expérimentées sont les suivantes :
\begin{itemize}
\item  8 métriques basées sur les caractères : JaroWinkler, Monge Elkan, Jaro, Levenshtein, Needleman Wunch, Smith Waterman, Qgram, I\_sub.
\item 2 métriques basées sur les tokens : Jaccard, TFIDF.
\item 3 métriques hybrides : JaroWinklerTFIDF (SoftTFIDF), TagLink, Monge Elkan hybride.
\item Nos 81 métriques : c'est-à-dire les métriques $Liuppa(i, j)$, $\forall i, j : 1 \leq i, j \leq 9$, combinant les métriques du tableau \ref{tableau_mu1_mu2_epsilon}.
\end{itemize}
Les méthodes ci-dessus (sauf les métriques $Liuppa(i, j)$) sont implémentées dans le projet Simmetrics (Levenshtein, Needleman Wunch, Smith Waterman, Qgram) et dans le projet SecondString (JaroWinkler, Monge Elkan, Jaro, TFIDF, JaroWinklerTFIDF (SoftTFIDF), TagLink, Monge Elkan 2). La méthode I\_sub \cite{StoilosSK05} nous a été communiquée directement par son auteur. Les résultats de l'expérimentation sur chaque jeu de données seront discutés ci-après.

\subsection{Expérimentation sur des données ontologiques}
\label{exp_sur_donnee_ontologique}

Le jeu de données est ici déterminé par des experts à partir de l'ontologie géographique du projet GéOnto. Chaque paire est composée d'un label des concepts de cette ontologie et d'un terme extrait du corpus représentant le qualifiant des toponymes. À l'aide d'experts nous avons défini 81 paires correctes et 48 paires incorrectes.
 Le tableau \ref{tableau_quelques_paires} illustre quelques paires correctes (=) et incorrectes (\#). La caractéristique principale de ce jeu de données est que l'ordre des tokens a de l'importance pour déterminer si des couples (terme, label) sont similaires, par exemple \textit{bureau de poste}  est différent de \textit{poste de radio}.

 \begin{table}[!h]
\subfigure{
	\includegraphics[width=1.0\textwidth]{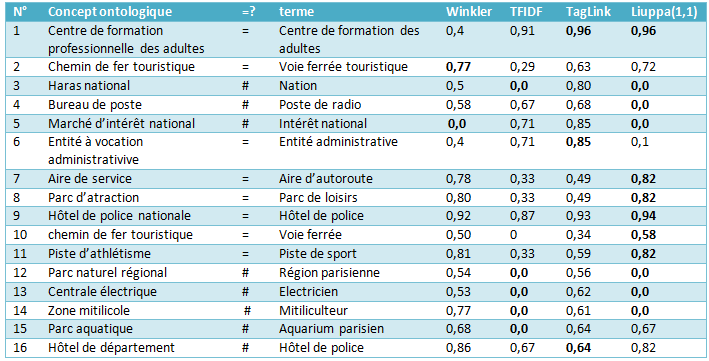}
}
\caption{Quelques paires et le score obtenu par différentes métriques}
\label{tableau_quelques_paires}
\end{table}


D'un point de vue quantitatif, le tableau \ref{tableau_performance_des_metriques} présente les métriques les plus performantes en fonction de la précision moyenne. L'expérimentation a montré que les 31 premières métriques sont celles produites par notre méta-méthode $Liuppa(i,j)$ et que la meilleure est la métrique $Liuppa(1,1)$. Cette métrique utilise JaroWinkler pour le niveau \og token\fg{} et pour le niveau \og séquence de symboles\fg{}. Cela peut être expliqué par le fait que JaroWinkler est une métrique bien adaptée à la comparaison de chaînes courtes \cite{Cohen2003} ce qui est généralement le cas des tokens (composés d'une dizaine de caractères au maximum) et des séquences de symboles (composées de moins de dix symboles) dans notre jeu de données.

\begin{table}[!h]
\centering
\subfigure{
	\includegraphics[width=0.4\textwidth]{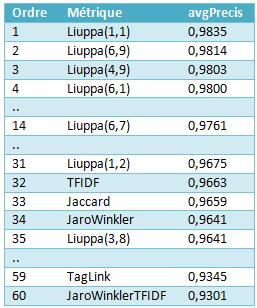}
}
\caption{Les meilleures métriques}
\label{tableau_performance_des_metriques}
\end{table}

D'un point de vue qualitatif, le tableau \ref{tableau_quelques_paires} présente des cas qui ne marchent pas pour les meilleures métriques dans chaque catégorie : JaroWinkler (basées sur caractères), TFIDF (basée sur token), TagLink(hybride) et notre métrique $Liuppa(1,1)$. En général, la comparaison caractère par caractère (JaroWinkler) n'est pas efficace si les chaînes sont longues (par exemple les paires 1, 6). Cependant la comparaison des tokens lexicalement proches peut présenter un intérêt. Par exemple, la paire~2 est correctement évaluée avec notre métrique, mais ne l'est pas avec TFIDF qui considère que \textit{fer}  est différent de \textit{ferrée}. Il existe cependant des effets de bord, puisque, pour la paire~15 notre métrique considère \textit{parc}  et \textit{parisien}  comme étant similaires tandis que cette paire est considérée comme différente par TFIDF. De la même manière pour la paire~16, les chaînes de caractères ont un préfixe en commun qui se compose de deux tokens (\textit{hôtel, de}), en conséquence, leur similarité est assez grande.

\subsection{Expérimentation sur des données issues de la campagne d'évaluation de Cohen}
\label{exp_sur_donnee_de_Cohen}
Pour cette deuxième expérimentation, nous avons repris le jeu de données de Cohen \cite{Cohen2003}. Dans ce jeu de données, une paire est composée de deux enregistrements de base de données qui partagent au moins un token ou un bloc de caractères. La paire est dite correcte si les enregistrements ont le même identifiant      par exemple \og White Ibis \fg{}, \og Ibis: White Ibis (Ibis blanc)  Eudocimus albus \fg{}  est une paire correcte). Dans le cas contraire, elle est dite incorrecte. La caractéristique principale de ce jeu de données est que l'ordre des tokens n'est pas important pour considérer que deux chaînes sont identiques (par exemple \og Ibis: Glossy \fg{}  est identique à \og Glossy Ibis \fg{}).

Le tableau \ref{tableau_experimentation_Cohen} présente quelques statistiques sur le jeu de données expérimenté en ce qui concerne le nombre de paires correctes et le nombre de paires incorrectes. L'expérimentation a montré que notre métrique est classée troisième parmi les meilleures métriques (figure \ref{figure_experimentation_Cohen}).
Notons que, en toute logique les deux meilleures métriques (TagLink, SoftTFIDF) sont ici celles qui ne prennent pas en compte l'ordre des mots dans des chaînes de caractères. 

\begin{table}[!h]
\centering
\scriptsize
\begin{tabular}{l | l | l}
fichier & nombre de paires correctes & nombre de paire incorrectes \\
\hline
birdScott1.txt & 15 & 5 \\
birdScott2.txt &  155 & 3785 \\
birdNybirdExtracted.txt & 55 & 2278 \\
birdKunkel.txt & 19 & 390 \\
business.txt & 295 & 165941 \\
census & 327 & 175979 \\
\end{tabular}
\caption{Le jeu de données de Cohen}
\label{tableau_experimentation_Cohen}
\end{table}

\begin{table}[!h]
\centering

\subfigure{
\centering
\includegraphics[width=0.4\textwidth]{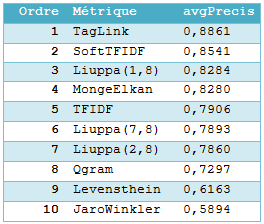}
}
\caption{Résultat d'expérimentation sur le jeu de données de Cohen}
\label{figure_experimentation_Cohen}
\end{table}

%

\bigskip
À partir de ces deux expérimentions, nous pouvons conclure que notre méthode est la meilleure lorsque l'ordre des mots a de l'importance,  mais comme elle est paramétrable elle reste intéressante avec d'autres types de jeu de données.

\section{Conclusion}
Dans cet article, nous avons proposé une méthode de comparaison de chaînes de caractères dont l'objectif final est de comparer des termes composés d'un ou plusieurs mots avec les labels des concepts d'une ontologie. Cette méthode est plus performante dans le cas où les éléments à comparer sont composées de plusieurs mots au sein desquels l'ordre a de l'importance.

La particularité de l'approche hybride proposée dans cet article est de combiner deux métriques basées sur les caractères (au lieu d'en utiliser une seule comme dans les autres approches hybrides) pour comparer les chaînes à deux niveaux différents : niveau des tokens et niveau des séquences de symboles. La combinaison deux à deux des 9 métriques de base produit 81 nouvelles métriques hybrides.
Dans le cadre d'une première évaluation, ces combinaisons ont été expérimentées sur un échantillon de paires de mots ou groupes de mots issues d'un vocabulaire lexical, d'une part, et des labels de concept  d'une ontologie, d'autre part.
Lors de cette expérimentation, nous proposons une démarche d'utilisation de la méta méthode $Liuppa(i,j)$ en deux temps :
\begin{enumerate}
\item Expérimenter sur un échantillon représentatif de paires pour déterminer le meilleur paramétrage.  L'expérimentation de cette étape a montré que, dans notre contexte, la métrique $Liuppa(JaroWinkler, JaroWinkler)$ avec le seuil $\varepsilon=0,84$ donne de meilleurs résultats que 13 autres métriques de la littérature ;
\item Appliquer la métrique déterminée à l'étape 1 à l'ensemble des paires à évaluer. En effet, nous avons intégré notre métrique $Liuppa(JaroWinkler, JaroWinkler)$ dans notre chaîne de traitement afin de comparer des termes extraits du texte avec les labels associés aux concepts de l'ontologie géographique de l'IGN.
\end{enumerate}

Dans un contexte de recherche d'information, la métrique $Liuppa(JaroWinkler, JaroWinkler)$ nous permet d'exploiter une ontologie géographique afin de typer des entités nommées spatiales extraites de récits de voyages. De plus, nous avons montré que la méta-méthode $Liuppa(i,j)$ proposée ici peut être paramétrée automatiquement à partir d'un échantillon expérimental pour être appliquée à d'autres types de données (comparaison d'enregistrements de bases de données, par exemple).

Dans le cadre de travaux relatifs à la recherche d'information sémantique, $Liuppa(i,j)$ pourra être expérimentée sur des domaines différents faisant appel à des ontologies dédiées et, éventuellement, de nouveaux corpus. Dans le contexte d'appariement d'ontologie, notre méthode peut être intégrée dans les outils dédiés à l'appariement d'ontologie tel que Taxomap \cite{Hamdi2008}, S-Match \cite{Giunchiglia2003}, etc. pour comparer les entités des ontologies au niveau élémentaire avant de les comparer au niveau structurel.
\vspace{-0,5cm}
\bibliography{biblio}
\label{end_document}
\end{document}